\documentclass{IEEEcsmag}

\usepackage[colorlinks,urlcolor=blue,linkcolor=blue,citecolor=blue]{hyperref}

\usepackage{upmath}

\usepackage{subcaption}
\usepackage[dvipsnames]{xcolor}
\definecolor{seccolor}{RGB}{150, 120, 20}
\usepackage{dblfloatfix}
\newcommand{\revise}[1]{\textcolor{black}{#1}}
\newcommand{\rmy}[1]{\textcolor{black}{#1}}

\usepackage{listings}
\usepackage{xcolor}

\definecolor{codegreen}{rgb}{0,0.6,0}
\definecolor{codegray}{rgb}{0.5,0.5,0.5}
\definecolor{codepurple}{rgb}{0.58,0,0.82}
\definecolor{backcolour}{rgb}{0.95,0.95,0.92}

\lstdefinestyle{mystyle}{
    backgroundcolor=\color{backcolour},   
    commentstyle=\color{codegreen},
    keywordstyle=\color{magenta},
    numberstyle=\tiny\color{codegray},
    stringstyle=\color{codepurple},
    basicstyle=\ttfamily\scriptsize,
    breakatwhitespace=false,         
    breaklines=true,                 
    captionpos=b,                    
    keepspaces=true,                 
    numbers=left,                    
    numbersep=5pt,                  
    showspaces=false,                
    showstringspaces=false,
    showtabs=false,                  
    tabsize=6
}

\jvol{XX}
\jnum{XX}
\paper{8}
\jmonth{October/November}
\jname{Computer}
\pubyear{2021}

\begin{document}


\title{Privacy Guarantees of BLE Contact Tracing: A Case Study on COVIDWISE\textsuperscript{\textsection}}

\author{Salman Ahmed}
\affil{Virginia Tech}

\author{Ya Xiao}
\affil{Virginia Tech}

\author{Taejoong (Tijay) Chung}
\affil{Virginia Tech}

\author{Carol Fung}
\affil{Virginia Commonwealth University}

\author{Moti Yung\textsuperscript{\textsection}}
\affil{Google LLC and Columbia University}

\author{Danfeng (Daphne) Yao}
\affil{Virginia Tech}


\begin{abstract}
 
Google and Apple jointly introduced a digital contact tracing technology and an API called ``exposure notification,'' to help health organizations and governments with contact tracing.  
The
technology and its interplay with security and privacy constraints require investigation. In this study, we examine and analyze the security, privacy, and reliability of the technology with actual and typical scenarios (and expected typical adversary in mind),
and quite realistic use cases. 
We do it in the context of Virginia's COVIDWISE app.
This experimental analysis validates the properties
of the system under the above conditions, a result that seems crucial for the peace of mind of the exposure notification technology adopting authorities, and may also help
with the system's transparency and overall user trust.
\end{abstract}

\maketitle
\begingroup\renewcommand\thefootnote{\textsection}
\footnotetext{© 2021 IEEE. Personal use of this material is permitted. Permission from IEEE must be obtained for all other uses, in any current or future media, including reprinting/republishing this material for advertising or promotional purposes, creating new collective works, for resale or redistribution to servers or lists, or reuse of any copyrighted component of this work in other works.}
\footnotetext{The opinions and statements in this work (performed as a project within an academic setting) are personal, and do not necessarily represent the employer of this author.}
\endgroup


\chapterinitial{COVID-19} has become the most deadly viral outbreak across the globe since the H1N1 virus pandemic of 1918 (known as ``The Spanish influenza'' due to a misconception about its origin). Today, containment and mitigation have been the best strategies, at the start, in the absence of vaccination, and then after initial vaccines have been found, as the strategy when new waves of variants and mutations of the virus appear.
Contact tracing can greatly help early containment by tracing from people exposed to newly infected patients and isolating them early~\cite{Ferretti}. The latest advancement in computer technology aids the contact tracing process by tracking individuals’ mobile devices and their proximity using Global Positioning Systems~\cite{reichert2020privacy}, or Bluetooth Low Energy (BLE) beacons~\cite{troncoso2020decentralized}. 



\rmy{To combat COVID-19 and aid governments and health organizations with contact tracing, technology companies (Google and Apple, in particular) jointly introduced a  Bluetooth Low Energy (BLE) technology called Google/Apple Exposure Notification (GAEN) system in April 2020~\cite{gaen2020}.} The GAEN system uses interoperable BLE signals to broadcast Bluetooth beacons from one device to another when Android/iOS users come in close proximity. The Bluetooth beacons help track the distance between the users and the duration of users being in close proximity. When one person is diagnosed as COVID-19 positive at the time of the contact or within a valid time frame of the contact (and only then) the system can notify the other users about potential exposure to a COVID-19 positive person (the infected user uploads the generators of its signal and other users pull them from the server).

\rmy{Researchers have scrutinized the contact tracing technology and warned that adoption of the technology can have privacy and security issues~\cite{gvili2020security, baumgartner2020mind, paparazzi2020, littlethumb2020, iovino2020effectiveness, gennaro2020exposure, boutet2020contact}, thus perhaps advocating against its wide adoption. However, these works primarily designed attacks based on abstract protocol design and abstract (theoretically formulated) adversaries (which have access to much more than the actual protocol on actual devices given to users). These attacks, at times, represent extreme settings and economically unjustified (i.e., expensive) scenarios, rather than a typical adversary whose motivation is based on rational decision making and whose means are of a typical user. Most did not verify the systems based on actual device investigation (accessing the software itself and experimenting with devices), and none of them try to find out in which scenarios the system is robust enough against a typical attack.}

\rmy{While scrutiny is always important, none of the earlier works assessed the  feasibility of the attacks in the real situation (when the system is deployed) in terms of operation or cost feasibility vs. gain.
%
Granted, there are extreme cases of heavy investing and massive deployments of devices/readers which can attack the system, or
attacks that are extreme due to the attacker assumed capability (having full access to attacked devices). In this sense, the above attacks were, indeed, good to know as  extreme but unlikely attacker behavior.
This work, like other works~\cite{park2020contextualizing, park2018structuration, shin2010effects, shin2010ubiquitous} that evaluate trust, security, privacy, usefulness, traceability, transparency, and reliability, means to fill the gap and investigate the systems in a balanced way, by inspecting the actual system (software and operation of the working system), and also by assessing strengths and not only weaknesses (essentially, assuming the system encounters a typical not too costly attack by an adversary exploiting it directly, and not an adversary performing a dedicated targeted costly attack, by an attacker who invests much money to subvert properties of a system).}

We then increase the capabilities of the adversary gradually.
The current investigation can be useful to understanding the mitigation capability of the system against typical attacks as an explanation toward the system adoption-- currently, in future waves of the COVID-19 pandemic, or future pandemic outbreaks.

\rmy{Specifically, we perform an analysis of GAEN with two focus points: i) ensuring that the library code (from Google and Apple) and contact tracing app code (from various government and 
health organizations) protect user privacy, and ii) investigating the privacy 
shortcomings/flaws in the design and implementation of GAEN, if any.}

\revise{We investigate the above in the context of COVIDWISE~\cite{covidwise-va}, the state of Virginia's official contact tracing app, which uses the GAEN system. Other major GAEN-based contact tracing apps around the globe include COVID Alert (Canada), Corona-Warn-App (Germany), COVID Tracker (Ireland), SwissCovid (Switzerland), Immuni (Italy), NHS COVID-19 (United Kingdom), as well as several US apps including GuideSafe (Alabama), Covid Watch (Arizona), COVID Alert NY (New York), Care19 Alert (Wyoming), Safer Illinois (University of Illinois) and PocketCare  (University at Buffalo).
In mid March 2021, COVIDWISE was  the most adopted (10.5\%) contact tracing app in the US~\cite{gaen-adoption}. 
} 


In the rest of this article, we explain and analyze GAEN's privacy design. We experimentally evaluate several BLE-related properties. We confirm that GAEN prevents tracking through random Bluetooth addresses, thus providing strong privacy guarantees. We found that iPhones deliver strong privacy protection via the non-resolvable random private address and prevent malicious apps from snooping on users' Rolling  Proximity Identifiers (RPIs). We also confirm that RPI's refreshing interval is within the range of 10-20 minutes~\cite{googleapplebtspec2020} and may vary with the distance between devices.  For advanced attacks targeting contact tracing apps, we break down their assumptions and assess the attack feasibility.

    



\section{DESIGN OVERVIEW of GAEN}
\begin{figure}[!h]
\centerline{\includegraphics[width=14.7pc]{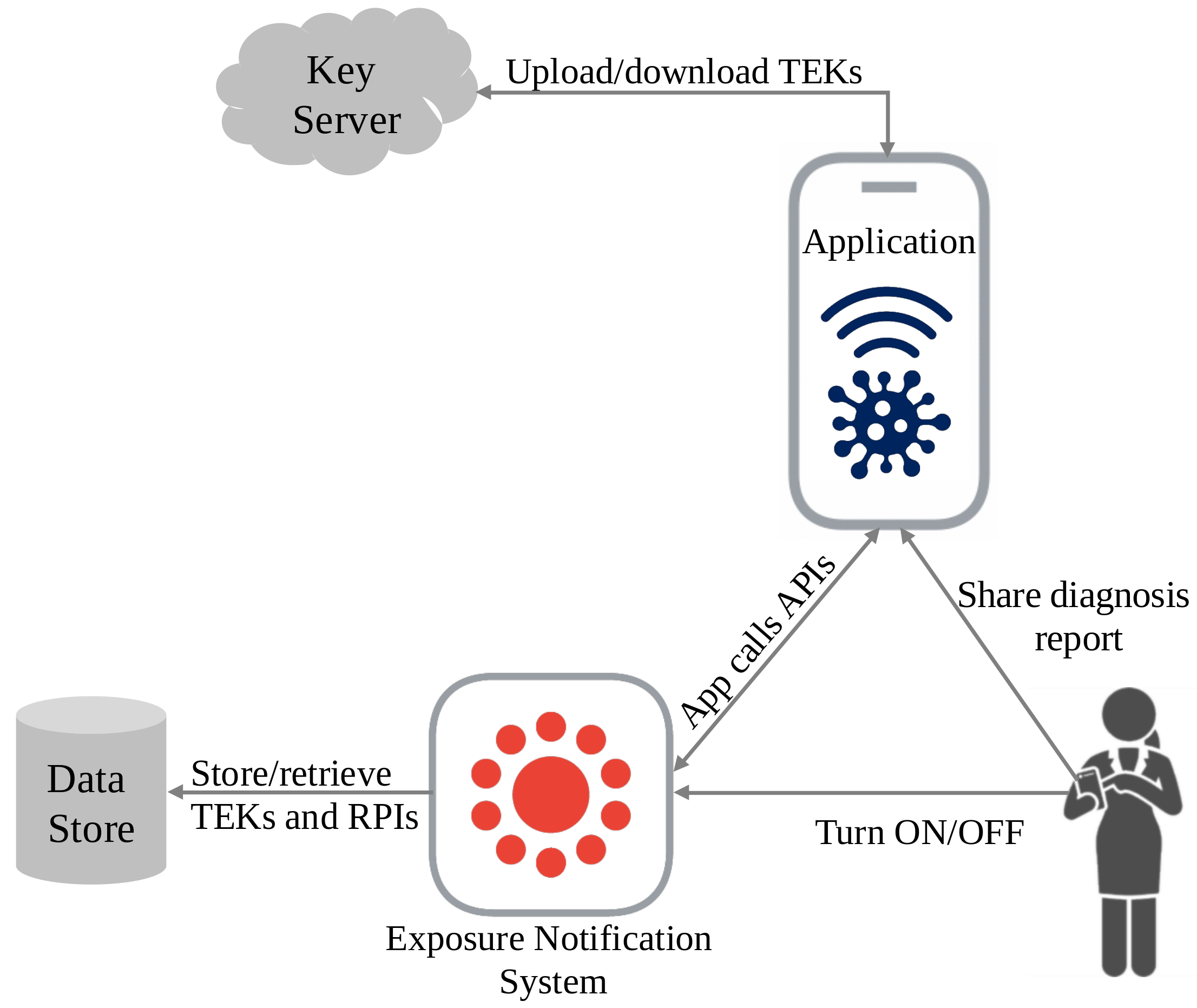}}
\caption{Interactions of user and contact tracing app with the exposure notification system.}
\label{gaen-diagram}
\end{figure}

\noindent
GAEN broadcasts and stores BLE beacons without any interaction with the app if the system is turned ON. However, a user can turn ON/OFF the system by either using the app or directly through the exposure notification settings. GAEN 
provides API calls to support different operations invoked by the contact tracing app. Figure~\ref{gaen-diagram} illustrates the interactions between a user, the exposure notification system, and the app.

GAEN uses BLE technology due to its availability on a smartphone 
needed for applications such as smart homes, proximity, wearables, healthcare, and fitness that require low data transfer and low latency. BLE's practical communication range is 10-20 meters (33-66 feet) which is sufficient for GAEN. GAEN's use of BLE also optimizes power consumption; it consumes significantly less power than traditional BLE communication used for peripherals. 
Besides, GAEN's passive power usage (i.e., actively broadcasting when devices are close to each other) further reduces power consumption.


\subsection{TEK, Bluetooth Beacon, and RPI}
\noindent
The heart of GAEN is a key called Temporary Exposure Key (TEK). TEK is a 16-byte random number generated using randomness and a cryptographically secure pseudorandom number generator. The TEK is associated with a device for a day within its lifetime. The GAEN system generates a new TEK every 24 hours to make it hard for attackers to track infected users, who employ and upload TEK numbers, beyond a day period. Then, the Bluetooth beacon’s payload includes an identifier called Rolling Proximity Identifier (RPI). An RPI is derived from a TEK as an AES encryption key
(and the current time indication within the 24 hours period as the message). In addition, a Bluetooth beacon’s payload also includes metadata such as protocol version and transmission power which are encrypted using a key derived from the TEK. The RPI and metadata are expected to change every 10-20 minutes (see the specification~\cite{googleapplebtspec2020}) to prevent attackers from tracking a device of uninfected users based on Bluetooth beacons being overly persistent.

%
When a user is infected, its TEKs for the relevant period (14 days) are uploaded to the server, and users pull TEKs of infected people from the server (not in the order they are uploaded), produce the day's RPIs, and match against their device stored RPI of that day and time,  done locally on the device to detect exposure. TEK being a daily key, makes it impossible to link RPIs between days (when one downloads TEKs from the server, there is no indication which TEK on other days are coming from the device of a given day's TEK). The goal of the system, from a basic privacy design goal, is to relate to TEKs and RPIs which are random objects and not to users or devices.
This design philosophy was originally shared by the GAEN system and many academic groups as well, all attempting to minimize the loss of privacy while allowing  support for contact tracing (and allowing reasonable storage and computations at devices).


\subsection{API and app responsibilities}
\noindent
The contact tracing app and the underlying GAEN system have different responsibilities. GAEN is responsible for transmitting, receiving, and storing Bluetooth signals. The app allows positively tested patients 
to share their diagnosis and automatically notify the central server and, eventually via the system, others who were in contact with the patients. The health authority (e.g., Virginia Department of Health) sets the exposure detection thresholds (e.g., the minimum distance between users and duration of exposure) used in the contact tracing app.

\begin{figure}[h!]
\centerline{\includegraphics[width=6.0pc]{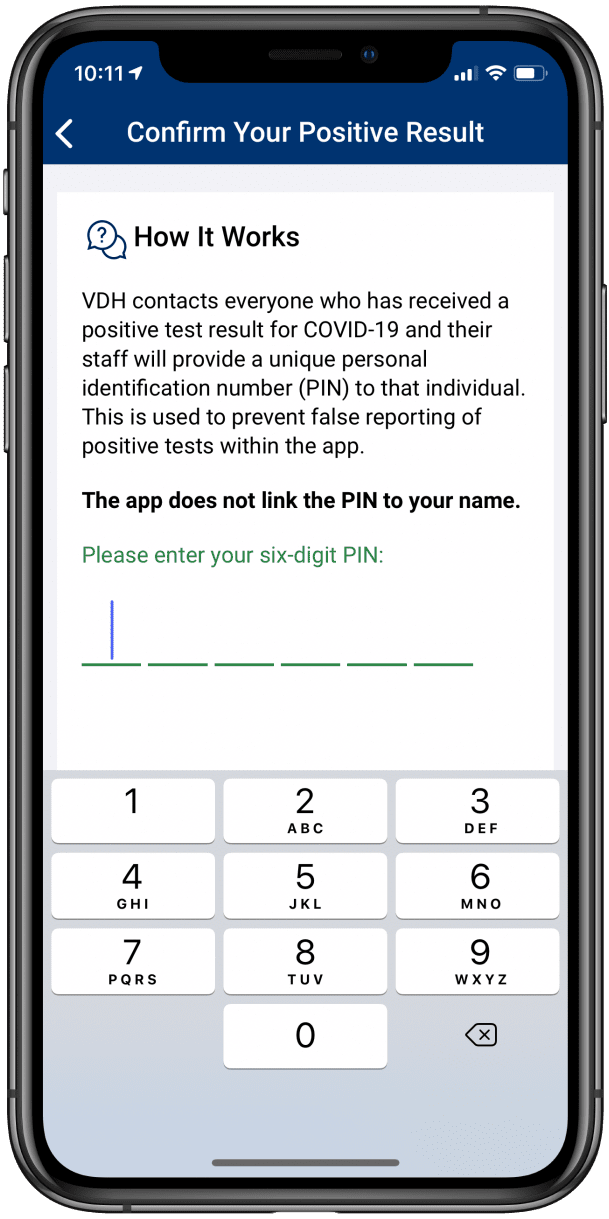}}
\caption{COVIDWISE allows a COVID-19 positive patient to share their positive information using a six-digit PIN.}
\label{covidwise-share}
\end{figure}


GAEN provides 17 API calls for a contact tracing app (e.g., COVIDWISE) to interact with the GAEN system. 
The key responsibilities of GAEN API calls are TEK creation and management, RPI generation and management, BLE broadcast and scan, and Exposure detection. The contact tracing app is responsible for user authorization, downloading published TEKs, presenting exposure notifications, and uploading TEKs if infected.
For example, the Virginia Department of Health, one of the health authorities using GAEN for their contract tracing app, allows users to share COVID-19 positive test results, and for security purposes, they assign a six-digit PIN to each COVID-19 patient.
The patient may enter the PIN to the COVIDWISE app (as shown in Figure~\ref{covidwise-share}).
This disclosure is voluntary in the State of Virginia.
People who have been in close proximity to a COVID-19 infected person for at least $T$ minutes in the last 14 days are then notified via the system operation on their device. The Virginia Department of Health determines and sets the value of $T$.


\subsection{Overview of GAEN's privacy design}
\noindent
Out of the 17 API calls that allow the interactions between the GAEN system and the app, two APIs – {\it getTemporaryExposureKeyHistory()} and {\it provideDiagnosisKeys()} in Android – deal with potentially sensitive information.
The first API call fetches the TEKs of the last 14 days from the on-device data store and provides them to the app for uploading to the key server. An app uses the second API call to insert one or more batches of TEKs into the on-device data store. These two API calls are sensitive because they exchange critical information (i.e., TEKs) between the app and the GAEN system. To ensure the privacy and integrity of the TEKs, the API calls use a specific file format (e.g., export.bin) and a verification method through signatures (e.g., export.sig).

On the other hand, the contact tracing app (COVIDWISE) is responsible for securely communicating with the key server, uploading, and downloading TEKs. Both the app and the key server verify the integrity of TEKs through digital signatures. The app  does not use any personally identifiable information (PII), device identifier, or Bluetooth identifier in the process of sharing the COVID-19 positive information. 

\subsection{Threat models and claimed privacy guarantees}
\noindent
We consider four threat levels to discuss GAEN's privacy guarantees: {\em i)} walking trail model, {\em ii)} your neighbor model, {\em iii)} stalker model, and {\em iv)} organized crime model. We define and categorize the threat levels based on attackers' privilege levels regarding accessing RPI beacons in different real-world scenarios. These privilege levels are also compatible with the assumptions made in the existing literature~\cite{gvili2020security, baumgartner2020mind, paparazzi2020, littlethumb2020, iovino2020effectiveness, gennaro2020exposure, boutet2020contact}. In the walking trail and your neighbor models, an adversary can sniff a very limited amount of beacons for obtaining RPIs. We then consider a stalking model where an adversary can sniff a small number of BLE beacons (e.g., using less than 10 BLE sniffing devices) to obtain RPIs. Finally, in an organized crime model, we assume that an adversary can compromise a smartphone, set up a large-scale infrastructure to sniff BLE beacons, and hack health care systems to obtain PINs to share positive information. We discuss the threat levels of these attack scenarios in Table~\ref{tab:privacy-risk-table}.

The privacy guarantees of GAEN and the contact tracing app lie in five key aspects: {\em i)} preventing attackers, public health authorities, government, and Apple/Google from tracking or monitoring a user’s movements, {\em ii)} generating TEKs without using any PII or any context (like the geographic location), {\em iii)} sharing COVID-19 positive diagnostic information without revealing any user information, {\em iv)} preventing attackers from obtaining any PII, even if attackers get access to the TEKs, and {\em v)} users’ ability to turn ON/OFF GAEN based on their discretion. Furthermore, based on the  principle of least privilege, TEKs never leave a user’s device unless the user tests positive.



\section{BLE and RPI EXPERIMENTS}
\noindent
We conducted simple experiments to investigate various BLE aspects in GAEN and COVIDWISE to confirm the privacy guarantees. Using PacketLogger (an extension to the Xcode Apple developer tool) in iOS and Bluetooth system logs in Android, we intercepted and collected Bluetooth beacons to investigate whether all the intervals work as expected and whether there exist any identifiers (e.g., resolvable Bluetooth address) in the Bluetooth beacons. We also inspected the device storage for keys and identifiers in Android (Pixel 4a) and iOS (iPhone 7) devices by measuring the number of Bluetooth beacons sent in 24 hours.


\subsection{Randomness of Bluetooth address}
\noindent
We examined the randomness of Bluetooth addresses used in transmitting Bluetooth beacons to observe if a receiving entity can resolve the sender's Bluetooth address. We observed that both Android and iOS utilize random addresses to conceal the identity of the sender while transmitting advertisement packets, as expected. 

Android and iOS apply different types of random addresses. Android phones use resolvable random private addresses, while iPhones use non-resolvable random private addresses in advertising packets. The difference is that Android devices allow trusted parties (e.g., paired devices) to resolve the resolvable random private addresses. However, both operating systems preserve privacy, assuming that any paired Bluetooth devices (e.g., the user's own AirPods) are trustworthy. 
%

It is important to know that contact tracing apps do not require location permission in the latest version of Android (i.e., Android 11). Older versions of Android apps require the location settings to be turned ON for the Bluetooth communication to work (a strange capability given that in this application no location information is used whatsoever!).

\subsection{RPI interception}
\begin{figure}[h!]
\centerline{\includegraphics[width=18pc]{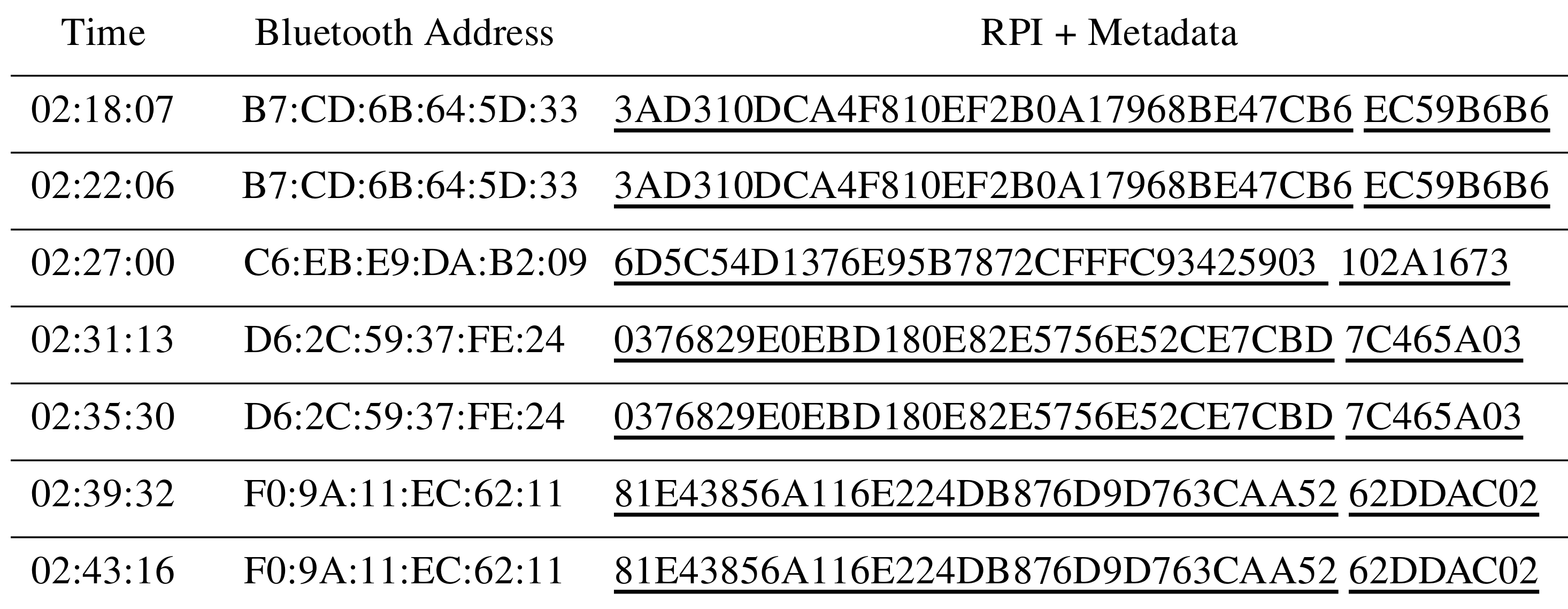}}
\caption{20-byte advertised random numbers with RPI and metadata captured from Pixel 4a. 
}
\label{raw-rpi-android}
\end{figure}

\noindent
We examined the runtime RPI (Bluetooth beacon) and metadata using PacketLogger for iOS devices and Bluetooth HCI snoop logs for Android devices. We observed that each device received a set of advertising payloads around every 4 minutes in Android and around every 3.5 minutes in iOS. Figure~\ref{raw-rpi-android} shows a raw BLE advertisement packet captured from an Android device (Pixel 4a). The last 20 bytes are composed of a 16-byte RPI and a 4-byte metadata.

We found that the RPI and metadata of the advising packets are zeros for iOS devices. The observed zero values are the result of third-party apps' access to the exposure notification UUID (Universally Unique Identifier). Because iOS blocks access to avoid malicious third-party apps from snooping on users' RPIs. 
This security mechanism renders attack proposals based on stealing RPIs (e.g., as described in~\cite{gaen-detector2020}) useless on iOS.

\begin{figure*}[h!]
\centerline{\includegraphics[width=37pc]{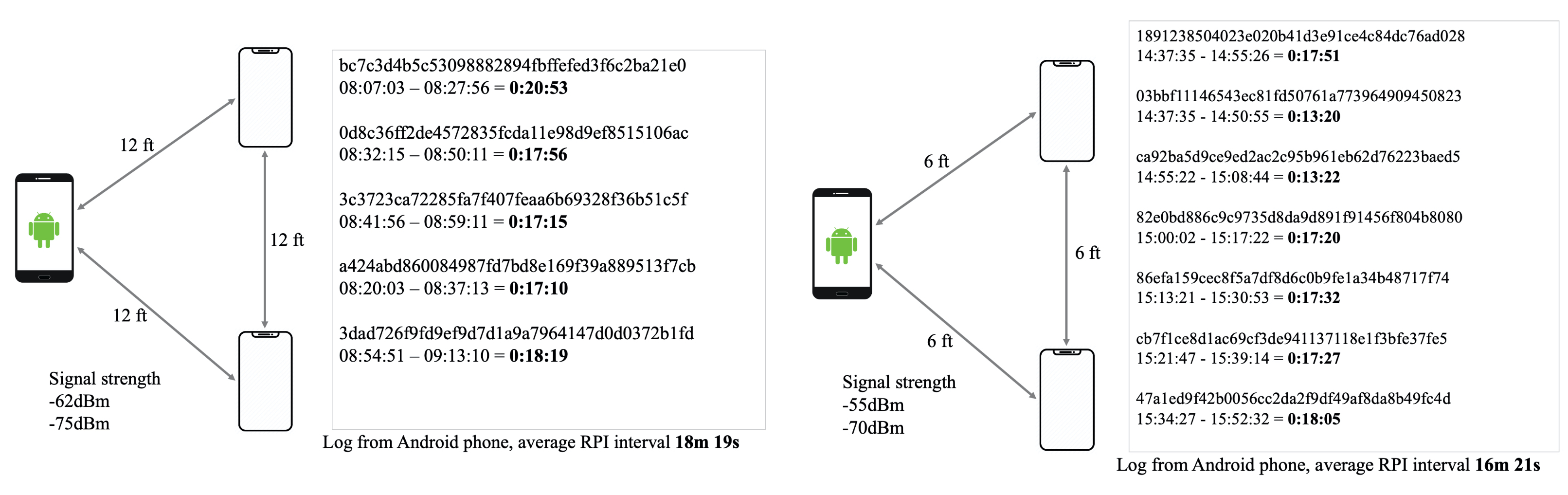}}
\caption{RPI transmission interval experiment. The interval varies with device distance.}
\label{rpi-tx-interval}
\end{figure*}

\subsection{RPI intervals}
\noindent
We intercepted Bluetooth beacons to examine the RPI transmission intervals. In our experiment, we used one Android device (Pixel 4a) and two iPhones (iPhone 7). 
We positioned each device according to the distance in Figure~\ref{rpi-tx-interval}. Based on RPIs received on the Android device, we observed that the RPI transmission interval varies with the distance between devices. However, the observed intervals satisfy the published intervals in the specifications (i.e., between 10 to 20 minutes). Imprecise distance approximation due to device positions (e.g., in human pockets) or surroundings (e.g., glass protections) may also impact this interval. 

\subsection{Key and RPI storage}
\noindent
We analyzed the storage requirements of RPIs. We found that an Android or iOS device takes around 0.59 MB to 0.63 MB of storage per day. Since GAEN stores keys and RPIs for the past 14 days, a device may take roughly 8.25 MB to 8.8 MB of storage if we consider non-interrupted interactions between the devices. In practice, the interactions are interrupted due to movements and obstacles, and the storage is less than 8 MB. 
For TEKs, we are unable to locate any stored TEK values on the logs of either Android or iOS devices, as expected. 



\section{GAEN's privacy w.r.t. threat models}
\noindent

As with all security solutions, the privacy guarantees of GAEN are relative. There certainly exist extreme scenarios (e.g.,~\cite{troncoso2020decentralized, baumgartner2020mind,  paparazzi2020,littlethumb2020, gaen-detector2020,chan2020pact}) where attackers may learn additional information. If an adversary has access to RPIs, TEKs, and RPI date-time information read by (say) thousands of users, then the adversary can  profile a user's movement~\cite{troncoso2020decentralized, paparazzi2020, baumgartner2020mind}. 
%

Table~\ref{tab:privacy-risk-table} summarizes the attack difficulty and the leak severity in GAEN under multiple (increasing) threat categories: {\em i)} walking trail causal encounter model, {\em ii)} your neighbor model, {\em iii)} stalker model, and {\em iv)} organized crime model. The first three models capture the most typical threat scenarios (representing small-scale individuals or group effort), in which, it turns out, GAEN leaks no sensitive information. 

The adversary in {\it Your Neighbor} model (ID 2) may occasionally   receive beacons from a few (e.g., 3-5) nearby users. The difference between {\it Your Neighbor} (ID 2) and {\it Stalker I} (ID 3) model is that adversaries in the former receive RPIs normally, while adversaries in {\it Stalker I} deliberately orient themselves (e.g., by changing locations) to intercept RPIs from more victims (e.g., 5-10). If successful, the {\it Stalker I} model only reveals the approximate number of nearby GAEN users, still not posing any privacy threat.


\begin{table*}[tb!]
	\centering
	\scriptsize
	\caption{Privacy leak and severity of leak in GAEN against realistic and complex threat models and their assumptions}
	\label{tab:privacy-risk-table}
	\begin{tabular}{|l|l|l|l|l|l|l|l|}
		\hline
		\hline
		ID & \begin{tabular}[c]{@{}l@{}}\textbf{Threat} \\ \textbf{Level}\end{tabular}  &\begin{tabular}[c]{@{}l@{}}\textbf{Attack} \\ \textbf{Difficulty}\end{tabular}  & \textbf{Attack Requirement} & \textbf{Attack Goal} & \textbf{Info Leaked} & \begin{tabular}[c]{@{}l@{}}\textbf{Severity} \\ \textbf{if Leak}\end{tabular} & \textbf{Refs} \\ \hline \hline
		1  & \begin{tabular}[c]{@{}l@{}}Walking \\Trail \end{tabular} & Low & \begin{tabular}[c]{@{}l@{}}Access to one RPI\\ (common scenario)\end{tabular} & \begin{tabular}[c]{@{}l@{}}Any information \\about a user\end{tabular} & None & None & \textemdash  \\ \hline
		2  &  \begin{tabular}[c]{@{}l@{}}Your \\Neighbor \end{tabular} & Low & \begin{tabular}[c]{@{}l@{}}Access to 0-5 RPIs from 3-5\\ victims considering neighbors\\ come nearby 0-5 times a day\\ (common scenario)\end{tabular} & \begin{tabular}[c]{@{}l@{}}Any information \\about a user\end{tabular} & None & None & \textemdash  \\ \hline
		3 & Stalker I & Low & \begin{tabular}[c]{@{}l@{}}Access to at least on RPIs in\\ a 10 to 20-minute time window \\from 5-10 victims\end{tabular} & \begin{tabular}[c]{@{}l@{}} To estimate the number \\ of GAEN users \\around an attacker\end{tabular} & \begin{tabular}[c]{@{}l@{}} Approximate number\\ of nearby GAEN users\end{tabular} & None & \cite{gaen-detector2020} \\ \hline
		4 & Stalker II & Medium & \begin{tabular}[c]{@{}l@{}}1. Access to RPIs from at \\ \ \ \ least one victim. Tracking \\ \ \ \ a victim for an hour \\ \ \ \ requires all RPIs in that hour\\ 2. Continuity of RPI\\ \ \ \ reception
			from a victim\end{tabular} & \begin{tabular}[c]{@{}l@{}} To continuously \\ track a user\end{tabular} & \begin{tabular}[c]{@{}l@{}} None (Not trackable \\ based on our \\observation)\end{tabular} & None & \cite{littlethumb2020}\\ \hline
		%
		5 & \begin{tabular}[c]{@{}l@{}}Organized \\Crime I \end{tabular} & High & \begin{tabular}[c]{@{}l@{}}1. Access to unlimited RPIs with \\ \ \ \  location data from 10+ victims\\ 2. Access to published \\ \ \ \ TEKs through jailbreaking \\  \ \ \ or rooting attacker’s phone or \\ \ \ \ imitating a contact tracing app\\3. Aggregated data for each \\ \ \ \ 10-20-minute time window:\\  \ \ \ date, time, interaction graph,\\  \ \ \ social graph, addresses,\\  \ \ \ location type (residential, \\ \ \ \ workplace, library, etc.),\\ \ \ \ surveillance cameras\end{tabular} & \begin{tabular}[c]{@{}l@{}}To profile movements\\
			of infected users \\and de-anonymize them\end{tabular} & \begin{tabular}[c]{@{}l@{}}Imprecise \\de-anonymization \\ (precision decreases \\with increasing \\ number of profiles\end{tabular} & Medium & \begin{tabular}[c]{@{}l@{}}\cite{troncoso2020decentralized},\\ \cite{baumgartner2020mind}, \\ \cite{paparazzi2020}\end{tabular} \\ \hline 
		6 & \begin{tabular}[c]{@{}l@{}}Organized \\Crime II \end{tabular} & High & \begin{tabular}[c]{@{}l@{}}1. Access to a victim’s \\ \ \ \ smartphone through hacking\\ 2. Storage protection bypass\end{tabular} & \begin{tabular}[c]{@{}l@{}}To obtain the victim’s\\infection status\end{tabular} & \begin{tabular}[c]{@{}l@{}}Information whether the \\victim is infected or not\end{tabular} & Medium & \cite{chan2020pact} \\ \hline
		\hline
	\end{tabular}
\end{table*}

A reported attack~\cite{littlethumb2020} relied on the asynchronous change of Bluetooth addresses and RPIs, which is represented in the {\it Stalker II} model (ID 4) in Table~\ref{tab:privacy-risk-table}. However, this attack no longer works, since GAEN requires the Bluetooth address and RPI  to change synchronously, which we experimentally confirmed by extracting around 11k random Bluetooth addresses and RPI pairs from the advertising packets over three days. We obtained advertising packets from an Android (v11) device (Pixel 4a), where the advertising packets were received from two iPhones (iPhone 7 and iPhone 11 with iOS 15.0.2) placing the three phones 6-12 feet away with regular phone activities. Each Bluetooth address is paired with a unique RPI and vice-versa. We checked the existence of any non-unique pairs by searching the usage of a Bluetooth address with multiple RPIs or an RPI with multiple Bluetooth addresses using a Python program. We observed no asynchronous change of the Bluetooth addresses and RPIs. Hence, user privacy is preserved in the {\it Stalker II} model (ID 4). Listing~\ref{tab:sync-change2} shows a few unique Bluetooth addresses and RPI pairs.

\lstset{
    escapeinside={(*}{*)}
}
\begin{lstlisting}[caption={Synchronous change of Bluetooth addresses and RPIs in advertising packets.}, label={tab:sync-change2}, basicstyle=\fontsize{5}{7}\ttfamily, numbers=none] 
    (*\bfseries Bluetooth Address*)       (*\bfseries Rolling Proximity Identifier (RPI)*)
    13:ac:57:35:3c:ea   59c62b86cdace1fe40446bc80689ccbd323588b8
    33:5d:64:6b:cd:b7   3ad310dca4f810ef2b0a17968be47cb6ec59b6b6
    09:b2:da:e9:eb:c6   6d5c54d1376e95b7872cfffc93425903102a1673
    24:fe:37:59:2c:d6   0376829e0ebd180e82e5756e52ce7cbd7c465a03
    04:2c:4d:b1:93:40   b5f1091b23a3871129a1225a6c3cebf175de28fa
\end{lstlisting}
Some attack scenarios in Table~\ref{tab:privacy-risk-table} have rather strong assumptions regarding the complexity of the attack setup and demand huge resources. For example, attackers in the {\it Organized Crime I} model (ID 5) require TEKs and aggregated data in each 10 to 20-minute time window to de-anonymize infected users~\cite{troncoso2020decentralized, baumgartner2020mind, paparazzi2020}. Aggregated data include public and sensitive information, such as date, time, interaction graph, social graph, address, location type (e.g., residential, workplace, and library), and surveillance cameras. This requirement of additional side-channel sources of information reduces the feasibility of the attack.  

In addition, the {\it Organized Crime I} model requires access to published TEKs through a jailbroken/rooted device or imitating a contact tracing app~\cite{baumgartner2020mind} (in normal operation, the TEK downloaded are not readily available
to the user and the exposure assessment is done away from the user). While obtaining TEKs through a jailbroken/rooted device might be feasible, imitating a contact tracing app is rather difficult. To imitate a contact tracing app, an attacker needs to somehow fool or bypass the authorization system, specifically an authorized administrative console, which is designed by GAEN to protect malicious apps from downloading TEKs. 

In addition, a malicious entity cannot fool the contact tracing app to accept forged TEK export files. For maintaining the back-end key server, an authorized contact tracing entity (e.g., Virginia Department of Health) must create a signing key to sign the TEK export files and share the corresponding public key with Google/Apple -- ensuring information authenticity.
%
%
%


Google and Apple also restrict app developers' access to GAEN APIs through an approval process. Google also limits the access to the Android Software Development Kit for regular app developers. These restrictions prevent the misuse and abuse of GAEN APIs.

The attack represented by the {\it Organized Crime II} model in Table~\ref{tab:privacy-risk-table} (ID 6) is difficult to launch in practice, as it requires the hacker to gain access to the victim's smartphone~\cite{chan2020pact}. 

While vulnerabilities like the power and storage drain do not hamper the effectiveness of GAEN,  vulnerabilities such as relay-and-replay and trolling attacks may degrade its effectiveness by increasing false positives. These false positives do not have an impact on privacy, though. Note that our reported results do not assess the effectiveness of GAEN, but rather they focus on privacy issues. (Hence, we do not discuss RPI spreading attacks which may require a large costly network of spreaders  and/or require large collaboration of smartphone holders which is unrealistic for a conspiracy to stay undetected: all these issues, while cheaper ``misinformation channels'' are available.)

\section{CONCLUSIONS}
\noindent
Our findings which are based, both, on experimentation with actual devices and a concrete system implementation, and on analysis based on classifying grades of attacks, confirmed that GAEN preserves privacy in a quite comprehensive collection of typical threat scenarios (including the walking trail casual encounter model, your neighbor model, the stalker model, and even in the costly organized crime model). Compromising user privacy by exploiting GAEN requires an unlikely,  complex, or  costly attack setup, e.g., compromising a victim's smartphone, mounting many Bluetooth radio devices, correlating with additional victim information, or rogue access to the healthcare systems. Besides, the built-in authorization, permission, and policy-enforcement mechanisms in GAEN add an extra layer of difficulty against the proposed attacks in the literature. 

To summarize, and in light of our findings, our article aims at helping people understand and appreciate GAEN's privacy protection, and encourage them to adopt GAEN-based contact tracing. This knowledge can be extremely powerful, as it will enable us to effectively manage the rest of the current (2020-2021) and future pandemics and, in turn, help reduce unnecessary casualties due to enhanced automatic contact tracing and its advantages, especially given the initial
estimates of effectiveness~\cite{Ferretti,Wymant}. 

\section{ACKNOWLEDGMENT}
This work has been supported by the Virginia Commonwealth Cyber Initiative (CCI).

\bibliographystyle{ieeetr}
\bibliography{ref}

\begin{IEEEbiography}{Salman Ahmed,}{\,} is a Ph.D. Candidate at Virginia Tech working with Professor Danfeng (Daphne) Yao. His research interests include system security assurance, attack surface quantification, and program analysis. He received his M.S. degree from East Tennessee State University. Contact him at ahmedms@vt.edu.

\end{IEEEbiography}

\begin{IEEEbiography}{Ya Xiao,}{\,}is a Ph.D. student in the Department of Computer Science at Virginia Tech, working with Professor Danfeng (Daphne) Yao. Her research interests include neural network-based software security solutions. 
She received an M.S. degree from Beijing University of Posts and Telecommunications (BUPT). Contact her at yax99@vt.edu.
\end{IEEEbiography}

\begin{IEEEbiography}{Dr. Taejoong (Tijay) Chung,}{\,} is an Assistant Professor at the Computer Science department at Virginia Tech. He received his Ph.D. in Computer Science and Engineering from Seoul National University in 2015. His work focuses on Internet security, privacy implications, and Internet measurement. 
Contact him at tijay@vt.edu.
\end{IEEEbiography}

\begin{IEEEbiography}{Dr. Carol Fung,}{\,} is an Associate Professor at Virginia Commonwealth University. She received her Ph.D. degree in computer science from the University of Waterloo (Canada). Her research interests include network security, mobile and IoT systems, and softwarized and programmable networks. 
Contact her at cfung@vcu.edu.
\end{IEEEbiography}

\begin{IEEEbiography}{Dr. Moti Yung,}{\,} is a Security and Privacy Research Scientist with Google and an Adjunct Research faculty with the Computer Science Department, Columbia University. He received his Ph.D. degree from Columbia University in 1988. Dr. Yung's research interests are primarily in security, privacy, and cryptography. 
Contact him at motiyung@gmail.com.
\end{IEEEbiography}
\begin{IEEEbiography}{Dr. Danfeng (Daphne) Yao,}{\,} is a Professor of Computer Science at Virginia Tech. She is an Elizabeth and James E. Turner Jr. ’56 Faculty Fellow and CACI Faculty Fellow. She received her Ph.D. degree from Brown University. Her research interests are on building deployable and proactive cyber defenses, focusing on detection accuracy and scalability. 
Contact her at danfeng@vt.edu.
\end{IEEEbiography}

\end{document}